\documentclass{bmvc2k}
\usepackage{multirow}
\usepackage{pifont}
\usepackage{amsmath}
\usepackage{amssymb}
\usepackage{multirow}
\usepackage{multicol}
\usepackage{float}
\pdfoutput=1

\title{QWD-GAN: Quality-aware Wavelet-driven GAN for Unsupervised Medical Microscopy Images Denoising}

\addauthor{Qijun Yang$^*$}{qijun.yang@manchester.ac.uk}{}
\addauthor{Yating Huang$^*$}{yating.huang@manchester.ac.uk}{}
\addauthor{Lintao Xiang}{ltxiang.work@gmail.com}{}
\addauthor{Hujun Yin$^\dagger$}{hujun.yin@manchester.ac.uk}{}

\addinstitution{
 Department of Electrical and Electronic
Engineering\\
 The University of Manchester\\
 Manchester, UK
}

\runninghead{Yang, et al.}{A Quality-aware Wavelet-driven GAN for Image Denoising}

\begin{document}

\renewcommand\thefootnote{}
\footnotetext{~\null\hspace{0pt}
  \noindent~\makebox[0pt][l]{%
    $^*$Equal contribution.\quad $^\dagger$Corresponding author.}
}
\renewcommand\thefootnote{\arabic{footnote}}

\maketitle

\begin{abstract}
Image denoising plays a critical role in biomedical and microscopy imaging, especially when acquiring wide-field fluorescence-stained images. This task faces challenges in multiple fronts, including limitations in image acquisition conditions, complex noise types, algorithm adaptability, and clinical application demands. Although many deep learning-based denoising techniques have demonstrated promising results, further improvements are needed in preserving image details, enhancing algorithmic efficiency, and increasing clinical interpretability. We propose an unsupervised image denoising method based on a Generative Adversarial Network (GAN) architecture. The approach introduces a multi-scale adaptive generator based on the Wavelet Transform and a dual-branch discriminator that integrates difference perception feature maps with original features. Experimental results on multiple biomedical microscopy image datasets show that the proposed model achieves state-of-the-art denoising performance, particularly excelling in the preservation of high-frequency information. Furthermore, the dual-branch discriminator is seamlessly compatible with various GAN frameworks. The proposed quality-aware, wavelet-driven GAN denoising model is termed as QWD-GAN.
\end{abstract}

\section{Introduction}
\label{sec:intro}

Biomedical microscopic imaging techniques such as fluorescence microscopy, confocal microscopy and two-photon microscopy play a crucial role in modern biomedical research and clinical diagnostics. However, these imaging techniques are inherently limited by physical constraints such as low photon counts, the need for rapid acquisition, and issues like photobleaching and phototoxicity during live-cell imaging, all of which make the imaging process highly susceptible to severe noise \cite{zhang2019poisson,zhou2020w2s}. As a result, the acquired images often exhibit complex noise patterns, including Poisson, Gaussian, and mixed Poisson-Gaussian distributions, posing significant challenges to accurate and reliable image analysis \cite{makitalo2012optimal,luisier2010image}.

\begin{figure*}[htbp]
\begin{center}
 \includegraphics[width=1.0\linewidth]{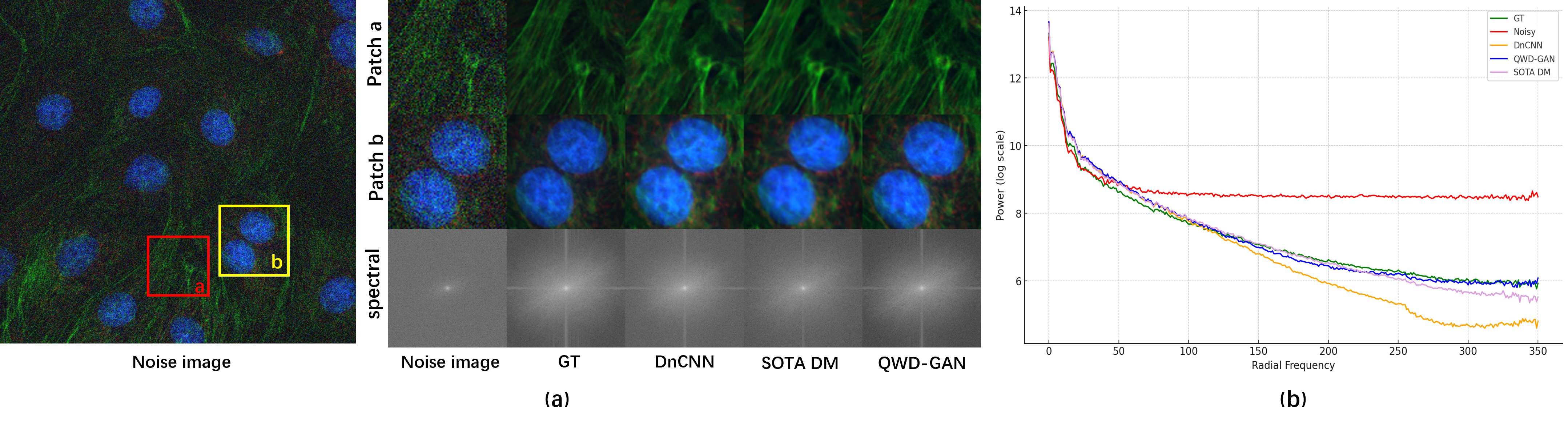}
\end{center}
\caption{Comparison of different noise denoising models on Fluorescence Microscopy Denoising (FMD) dataset \cite{zhang2019poisson}. Comparison of denoising of QWD-GAN in spatial domain and frequency domain (a), Radial frequency and power distribution curve (b).}
\label{fig:1}
\end{figure*}

When applied to real-world biomedical images, particularly under low signal-to-noise ratio (SNR) conditions, traditional denoising methods such as BM3D \cite{dabov2007image} and Pure-LET \cite{foi2010image} offer limited performance. Although deep learning-based approaches like DnCNN \cite{zhang2017beyond}, Noise2Noise \cite{lehtinen2018noise2noise}, and diffusion models have shown promising results, they often suffer from over-smoothing effects, as illustrated in Figure \ref{fig:1}. This can lead to the loss of structural details in high-frequency regions and dramatic performance degradation under varying imaging conditions, potentially resulting in blurred boundaries, nuclear membranes, microfilaments, and organelles. For example, although diffusion models demonstrate great potential in image denoising, they can generate hallucinated structures such as false cells or inaccurate organ boundaries under high level of noise or insufficiently trained conditions, which undermine diagnostic reliability \cite{pfaff2024no,liu2024residual}. Moreover, most existing diffusion models do not consider specific noise types or frequency characteristics unique to microscopic images, such as optical distortions and sample-induced diffraction blurs \cite{zhang2019poisson,makitalo2012optimal,chen2021residual}. 

Motivated by these challenges, we focus on developing a deep learning-based denoising model with a large receptive field to emphasize fine details such as edges, textures, and structures in microscopic images. In this work, we adopt a GAN-based framework and design a novel generator that integrates wavelet convolutions and residual convolutions. The generator fuses features from both branches using a frequency-attentive difference-aware module in a multi-scale setting. To further enhance the clinical applicability and interpretability of the generated images, we propose a dual-branch discriminator that incorporates image quality feature maps alongside original features, enabling the discriminator to better understand visual quality rather than relying solely on texture adversarial cues.
Our main contributions are summarized as follows:

(1) A wavelet-guided multi-scale generator is proposed that fuses spatial and frequency-domain features via wavelet and residual convolutions, effectively enhancing the preservation of high-frequency details in microscopy images.

(2) A quality-aware dual-branch discriminator is introduced that incorporates perceptual image quality features to better distinguish structural artifacts and guide the generator toward clinically reliable outputs.

(3) We define a new evaluation metric, High-Frequency Retention Ratio (HFRR), to measure the fidelity of fine structures, and demonstrate through extensive experiments on the FMD \cite{zhang2019poisson}, W2S \cite{zhou2020w2s}, and REFUGE \cite{orlando2020refuge} datasets that our model achieves state-of-the-art denoising performance with strong for generalization ability.

\section{Related Work}

\textbf{Biomedical Image Denoising.} Biomedical imaging, such as fluorescence microscopy and ultrasound, often suffers from photon-limited and acquisition-related noise. Classical priors like Non-Local Means (NLM) \cite{buades2005nonlocal}, BM3D \cite{dabov2007image}, and WNNM \cite{gu2014wnnm} exploit patch redundancy or low-rank assumptions but underperform on complex biological structures. Non-learning based image denoisers have tried to reconstruct clean images using pre-defined priors which model the distribution of noise \cite{mairal2009non,osher2005iterative,xu2007iterative,xu2018trilateral,zoran2011learning,xu2017multi}. Deep learning models such as DnCNN \cite{zhang2017beyond}, FFDNet \cite{zhang2018ffdnet}, and MemNet \cite{tai2017memnet} improve upon these by learning end-to-end mappings. More recent transformers like Uformer \cite{wang2022uformer} and Restormer \cite{zamir2022restormer} leverage self-attention for long-range dependencies. DeamNet \cite{ren2021adaptive} further introduces an adaptive consistency prior for robust denoising. However, supervised methods require clean/noisy pairs, prompting unsupervised solutions such as Noise2Noise \cite{lehtinen2018noise2noise}, Noise2Void \cite{krull2019noise2void}. Recent years, Diffusion-based denoising methods such as DDPM \cite{ho2020denoising}, DDIM \cite{song2020denoising} and RDDM \cite{liu2024residual} recast denoising as a generative sampling problem, where the clean image is progressively recovered from noisy observations via a reverse diffusion process tailored to the noise model, enabling flexible and high-fidelity restoration across various degradation types \cite{saharia2022image,lugmayr2022repaint}. However, Laura et al. pointed out that diffusion models may remove or introduce unrealistic structures, affecting the interpretability and reliability of clinical diagnosis \cite{pfaff2024no}.

\noindent \textbf{Multi-Scale Architectures.} Multi-scale architectures effectively aggregate fine- and coarse-grained information for denoising. Encoder-decoder designs (e.g. U-Net) and parallel-branch models like HRNet \cite{sun2019hrnet}, PANet \cite{zhang2019panet}, and MSCNN \cite{ren2020single} fuse multi-resolution features. SADNet \cite{tai2017memnet} and CLEARER \cite{wu2021clearer} enhance this with spatial-adaptive or search-based fusion strategies. MSANet \cite{lee2024msanet} improves upon these by introducing scale-specific subnetworks to better capture within-scale statistics and cross-scale dependencies. MSA-Net \cite{gou2022msanet} complements this idea using multi-scale self-attention to leverage non-local correlations across resolutions, achieving strong performance in biomedical denoising tasks.

\noindent \textbf{Frequency Domain in CNNs.} Frequency-domain analysis provides complementary insights to spatial-domain features, especially since noise typically disturbs high-frequency components. Several studies integrate frequency-domain priors into CNNs. Xu et al. \cite{xu2020dct} accelerate training using Discrete Cosine Transform (DCT) basis. Dzanic et al. \cite{dzanic2020fourier} and Durall et al. \cite{durall2020deepfake} show GANs produce unrealistic high-frequency statistics, motivating spectral regularization. Cai et al. \cite{cai2021frequencygan} propose frequency-aware discriminators, while Yang \& Soatto \cite{yang2020fda} utilize frequency alignment for domain adaptation. Inspired by these, Kim et al. \cite{kim2021uidfdk} proposed an unsupervised denoising method that introduces a spectral discriminator and frequency reconstruction loss to align high-frequency spectra between clean and generated images, achieving competitive results without paired data.

\section{Method}

\subsection{Wavelet-Guided Multi-Scale Denoising Generator}

\begin{figure*}[htbp]
\begin{center}
 \includegraphics[width=0.7\linewidth]{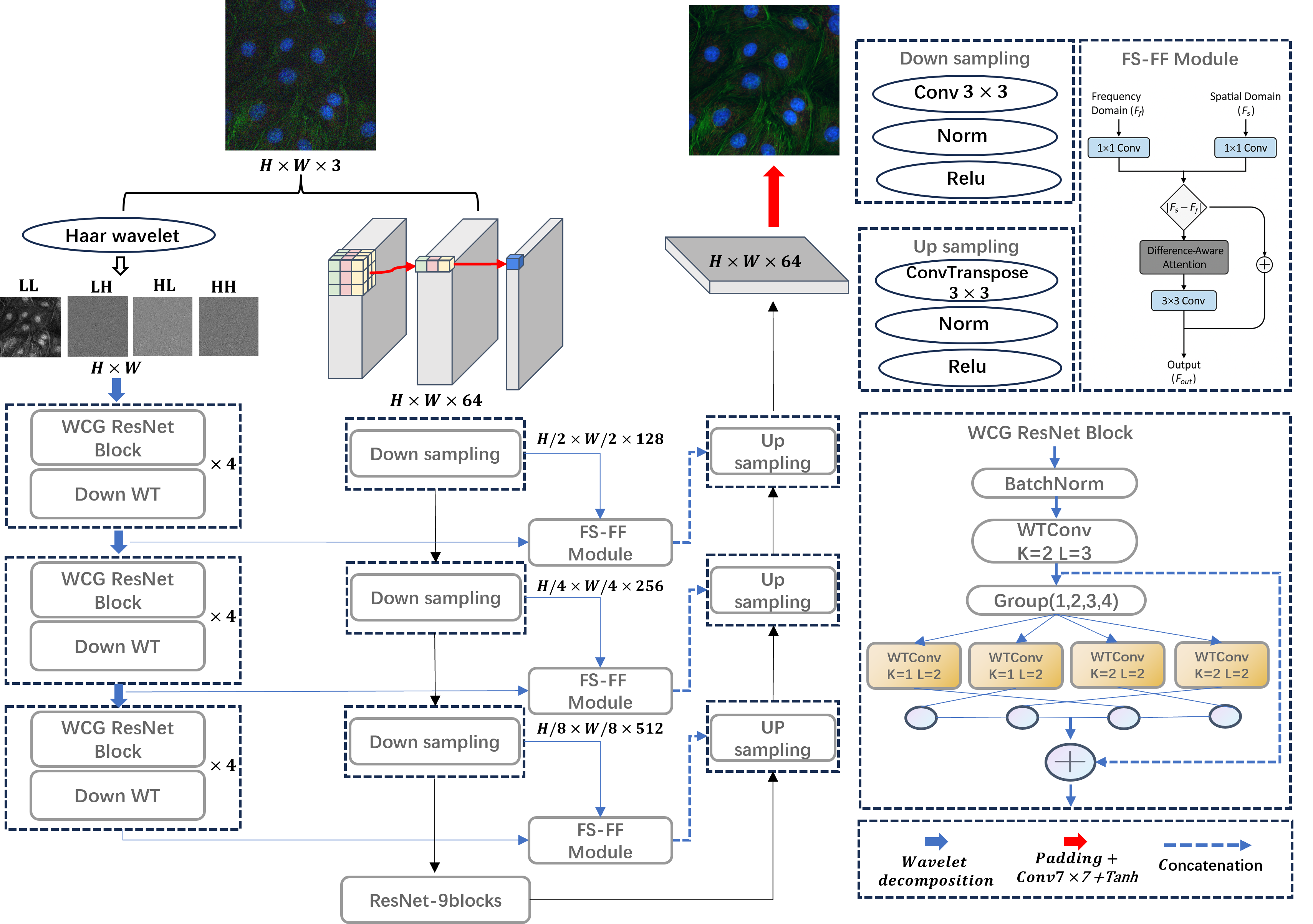}
\end{center}
   \caption{Structure of Wavelet-Guided Multi-Scale Denoising Generator}
\label{fig:2}
\end{figure*}

The Wavelet-Guided Multi-Scale Denoising Generator (WG-MSDG) is the core component of the proposed QWD-GAN architecture which is shown in Figure \ref{fig:2}. It is specifically designed to address the challenge of denoising biomedical microscopy images while preserving fine structural details such as organelle boundaries, microfilaments, and membrane edges. The generator achieves this by synergistically combining wavelet-domain frequency analysis with spatial-domain feature encoding in a multi-scale architecture.

\noindent \textbf{Architecture Overview.} WG-MSDG consists of two main branches: (1) a \textit{wavelet convolution branch} that processes frequency-domain components extracted via wavelet decomposition, and (2) a \textit{residual convolution branch} that models spatial information through standard CNN blocks. The outputs of both branches are merged by a Frequency-Spatial Feature Fusion (FS-FF) module to form a unified multi-scale representation. This design allows the model to leverage both fine textures and semantic structures across different resolutions. The specific design details and relevant formulas of Wavelet-Convolutional Group (WCG) Block and FS-FF module are shown in the supplementary material.

\noindent \textbf{Wavelet-Convolutional Grouped ResNet.} Let $x \in \mathbb{R}^{C \times H \times W}$ be an input feature map. We first decompose $x$ into $G$ channel groups $\{x_g\}_{g=1}^{G}$, and apply grouped wavelet convolutions $\text{WTC}_g$ to each:

\begin{equation}
W(x) = \bigoplus_{g=1}^{G} \text{WTC}_g(x_g)
\end{equation}

Each $\text{WTC}_g$ operates on a different group of features using learnable kernels in the wavelet domain. The output of the wavelet convolutional block is combined with the input via a residual connection:

\begin{equation}
y = x + \sum_{g=1}^{G} \text{WTC}_g(\text{BN}(x_g))
\end{equation}

This wavelet-convolutional grouped ResNet block (WCG-ResNet) improves the model's ability to encode high-frequency features while maintaining spatial coherence.

\noindent \textbf{Frequency-Spatial Feature Fusion (FS-FF).} The FS-FF module is responsible for aligning and integrating the features extracted from the wavelet and residual branches. It includes feature projection layers, difference-aware attention mechanisms, and multi-scale concatenation operations to enable cross-domain fusion. This module improves the complementarity of spatial and frequency information, enabling robust reconstruction under varying noise conditions.



\noindent \textbf{Loss Functions.} To optimize the generator, we employ a composite objective function defined as:

\begin{equation}
\mathcal{L}_{\text{total}} = \lambda_1 \mathcal{L}_{\text{recon}} + \lambda_2 \mathcal{L}_{\text{percep}} + \lambda_3 \mathcal{L}_{\text{wavelet}},
\label{eq:3}
\end{equation}
where:
\begin{itemize}
    \item \textbf{Pixel-wise Reconstruction Loss:}
    \begin{equation}
    \mathcal{L}_{\text{recon}} = \| \hat{y} - y \|_1,
    \end{equation}
    which measures the $L_1$ distance between the predicted image $\hat{y}$ and the ground truth image $y$.

    \item \textbf{Perceptual Loss:}
    \begin{equation}
    \mathcal{L}_{\text{percep}} = \sum_i \| \phi_i(\hat{y}) - \phi_i(y) \|_2,
    \end{equation}
    where $\phi_i(\cdot)$ denotes the feature map extracted from the $i$-th layer of a pre-trained VGG network. This term encourages perceptual similarity between $\hat{y}$ and $y$.

    \item \textbf{Wavelet Consistency Loss:}
    \begin{equation}
    \mathcal{L}_{\text{wavelet}} = \sum_{j,k} \| W_{j,k}(\hat{y}) - W_{j,k}(y) \|_1,
    \end{equation}
    where $W_{j,k}(\cdot)$ denotes the wavelet coefficients at scale $j$ and orientation $k$. This term enforces consistency in the wavelet domain. $\lambda_1$, $\lambda_2$, and $\lambda_3$ are weighting factors set empirically.
\end{itemize}

\noindent \textbf{Benefits.} The WG-MSDG architecture provides several benefits: (1) enhanced high-frequency detail preservation, (2) improved noise robustness across multiple scales, and (3) strong generalization for diverse biomedical imaging modalities. By embedding wavelet-domain processing directly into the generation pathway, our model effectively bridges spatial and frequency domains for superior denoising performance.

\subsection{Image quality-aware dual-branch discriminator}

\begin{figure*}[htbp]
\begin{center}
 \includegraphics[width=0.8\linewidth]{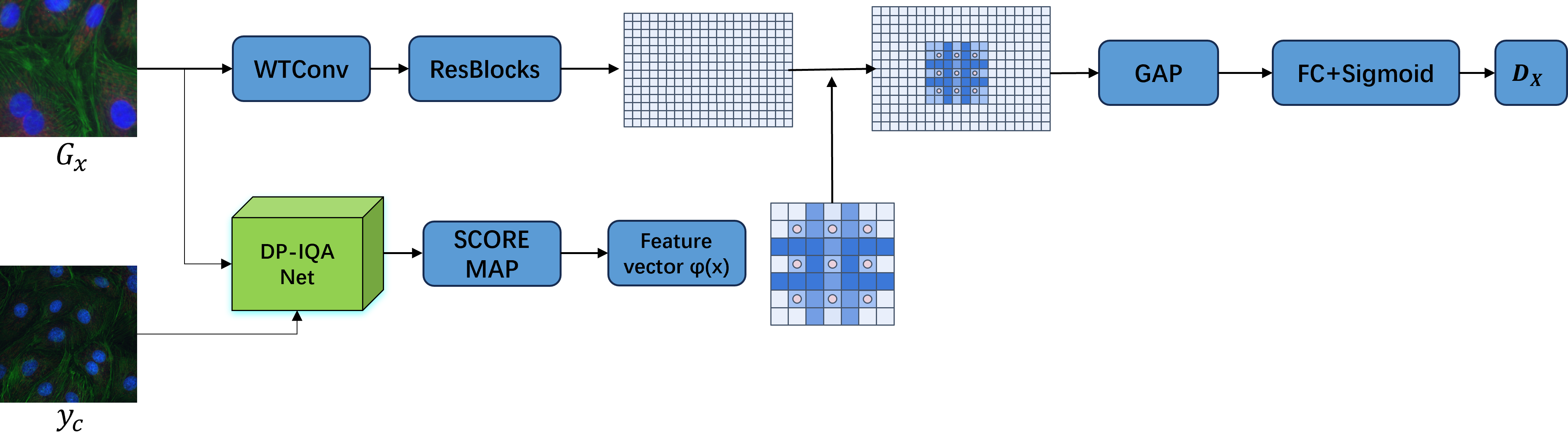}
\end{center}
\vspace{-0.5em}
   \caption{structure of Quality-Aware Discriminator}
\label{fig:3}
\end{figure*}

Although the proposed WG-MSDG generator is capable of producing visually clean and structurally detailed images, ensuring perceptual realism and clinical reliability requires more than pixel-wise or frequency-domain similarity.  Therefore, we propose an Image quality-aware dual-branch discriminator to guide the generator to produce output images with natural appearance and realistic structure. Based on the conventional convolutional backbone, this discriminator incorporates high-level features extracted from an Image Quality Assessment (IQA) module as auxiliary information, forming a perception-enhanced discrimination framework as shown in Figure \ref{fig:3}. The specific design details and relevant formulas for the Difference Perceptual Image Quality Assessment (DP-IQA) are shown in the supplementary material.

\noindent \textbf{Overall Architecture.} The input image is first processed through two parallel branches. The main branch applies standard convolutional layers and residual blocks to extract spatial features $F \in \mathbb{R}^{C \times H \times W}$. IQA branch passes the image through a pre-trained Image Quality Assessment (IQA) network (e.g., DP-IQA) to extract perceptual features $\phi(x)$, such as entropy maps or quality embeddings.

These two branches are fused using a \textit{spatial modulation} mechanism. If $\phi(x)$ is a normalized entropy map $\in \mathbb{R}^{1 \times H \times W}$, the fused features $F'$ are obtained as:

\begin{equation}
F'_{c,i,j} = F_{c,i,j} \cdot \frac{\phi(i,j)}{\max(\phi)}
\end{equation}

\noindent \textbf{Final Classification.} The fused feature map $F'$ is aggregated via global average pooling and passed through a fully connected layer followed by a sigmoid activation to produce the final adversarial score:

\begin{equation}
D(x) = \sigma(w^\top z + b), \quad z = \frac{1}{H \cdot W} \sum_{i,j} F'_{:,i,j}
\end{equation}

\noindent \textbf{Quality-Aware Loss.}
To further enhance perceptual consistency, we introduce a quality-aware loss for the generator:

\begin{equation}
\mathcal{L}_{\text{IQA}} = \sum_{l} \| \phi_l(x_{\text{real}}) - \phi_l(x_{\text{fake}}) \|_2^2
\end{equation}

This loss encourages the generator to produce images that match the perceptual distribution of real images. The final generator objective is:

\begin{equation}
\mathcal{L}_{G} = \mathcal{L}_{\text{GAN}} + \lambda_1 \mathcal{L}_{\text{L1}} + \lambda_2 \mathcal{L}_{\text{IQA}}
\label{eq:7}
\end{equation}

\section{Experiments}
\subsection{Implementation Details}

\noindent \textbf{Dataset.} To evaluate the denoising performance of our proposed method on medical images with large field of view and low lighting conditions, we utilized three datasets: the Fluorescence Microscopy Denoising (FMD) dataset \cite{zhang2019poisson}, the Widefield Microscopy (W2S) dataset \cite{zhou2020w2s} and the Fundus dataset REFUGE \cite{orlando2020refuge}. Noise levels were simulated by averaging 2, 4, 8, or 16 noisy frames, with ground truth images obtained by averaging 50 frames, and all images were cropped to 256×256 pixels.

\noindent \textbf{Experimental Settings.} We implement our method using PyTorch \cite{paszke2017automatic}, and conduct all experiments on NVIDIA GeForce RTX 4090 GPUs with 24 GB of memory. The Adam optimizer \cite{kingma2014adam} is used with a learning rate initialized to 0.0001. The loss weights are set to $\lambda_1 = 1.0$, $\lambda_2 = 0.1$, and $\lambda_3 = 0.5$ in Eq.\eqref{eq:3} and $\lambda_1 = 100$, $\lambda_2 = 15$ in Eq.\eqref{eq:7}. A batch size of 16 is used for all experiments. For synthetic noise removal, we randomly crop input patches of size $128 \times 128$, while for real-world noise removal, we use input patches of size $256 \times 256$.

\subsection{Comparison with State-of-the-arts}

In this section, we compare the proposed method with seven state-of-the-art representative methods, including traditional non-learning based methods, e.g. BM3D \cite{yang2018bm3dnet}, supervised learning methods, e,g. DnCNN \cite{zhang2017beyond} and RIDNet \cite{anwar2019real}, unsupervised methods such as N2N \cite{lehtinen2018noise2noise}, N2V \cite{krull2019noise2void} and diffusion model RDDM \cite{liu2024residual}. Based on the image-to-image translation benchmark \cite{liu2022bci,christiansen2018silico}, we use the structural similarity index metric (SSIM) and peak signal-to-noise ratio (PSNR) as evaluation metrics. In addition, we introduce the High-Frequency Retention Ratio (HFRR) as an evaluation metric. The design and derived formulas of HFRR are presented in the supplementary materials.

\begin{table}[htbp]
    \centering
    \label{tab:comparison-metrics W2S}
    \renewcommand{\arraystretch}{1.1}
    \setlength{\tabcolsep}{2pt}
    \small

    \resizebox{\textwidth}{!}{
    \begin{tabular}{l|cccc|c|cccc|c|cccc|c}
        \hline
        \multirow{2}{*}{\textbf{Methods}} & \multicolumn{4}{c|}{\textbf{SSIM ↑}} & \textbf{Avg} & \multicolumn{4}{c|}{\textbf{HFRR ($\rightarrow 1$)}} & \textbf{Avg} & \multicolumn{4}{c|}{\textbf{PSNR ↑ (dB)}} & \textbf{Avg} \\
        \cline{2-5} \cline{7-10} \cline{12-15}
         & 1 & 4 & 8 & 16 & & 1 & 4 & 8 & 16 & &1 & 4 & 8 & 16 & \\
        \hline
        BM3D\cite{yang2018bm3dnet}     & 0.808 & 0.883 & 0.907 & 0.929 & 0.876 & 1.198 & 1.176 & 1.159 & 1.134 & 1.167  & 24.423 & 29.564 & 33.985 & 36.924 & 31.224 \\
        DnCNN\cite{zhang2017beyond}    & 0.906 & 0.926 &0.946 & 0.972 & 0.937 & 0.766 & 0.794 & 0.815 & 0.829 & 0.801 & 30.531 & 33.533 & 36.212 & 38.268 & 34.636 \\
        RIDNet\cite{anwar2019real}   &0.896 & 0.905 & 0.926 & 0.942 & 0.917 & 0.709 & 0.754 & 0.795 & 0.809 & 0.767 & 30.321 & 33.343 & 36.014 & 38.568 & 34.561 \\
        N2N\cite{lehtinen2018noise2noise}  &0.919 & 0.923 & 0.948 & 0.960 & 0.937 & 0.849 & 0.863 & 0.886 & 0.901 & 0.874 & 32.405 & 36.404 & 37.596 & 38.435 & 36.210 \\
        N2V\cite{krull2019noise2void}    & 0.920 & 0.933 & 0.941 & 0.956 & 0.937 & 0.821 & 0.856 & 0.923 & 0.942 & 0.885 & 32.156 & 35.281 & 37.359 & 40.986 & 36.695 \\
        RDDM\cite{liu2024residual}    & \textbf{0.938} & \textbf{0.949} & 0.953 & 0.961& 0.950 & 0.893 & 0.903 & 0.915 & 0.919 & 0.907 & \textbf{33.452} & 36.223 & 38.059 & 40.556 & 37.072 \\
        \textbf{Ours} & 0.936 & 0.945 & \textbf{0.964} & \textbf{0.978} & \textbf{0.956} & \textbf{0.951} & \textbf{0.962} & \textbf{0.971} & \textbf{0.986} & \textbf{0.969} & 33.311 & \textbf{36.749} & \textbf{38.145} &\textbf{41.561}  & \textbf{37.191} \\
        \hline
    \end{tabular}
    }
    \caption{Quantitative comparisons with other methods in terms of the SSIM, HFRR and PSNR on the FMD dataset\cite{zhang2019poisson}.}
  \label{Tab:1}
\end{table}

\begin{table}[htbp]
    \centering
    \label{tab:comparison-metrics-W2S-2}
    \renewcommand{\arraystretch}{1.1}
    \setlength{\tabcolsep}{2pt}
    \small
    \resizebox{\textwidth}{!}{
    \begin{tabular}{l|cccc|c|cccc|c|cccc|c}
        \hline
        \multirow{2}{*}{\textbf{Methods}} & \multicolumn{4}{c|}{\textbf{SSIM ↑}} & \textbf{Avg} & \multicolumn{4}{c|}{\textbf{HFRR ($\rightarrow 1$)}} & \textbf{Avg} & \multicolumn{4}{c|}{\textbf{PSNR ↑ (dB)}} & \textbf{Avg} \\
        \cline{2-5} \cline{7-10} \cline{12-15}
         & 1 & 4 & 8 & 16 & & 1 & 4 & 8 & 16 & &1 & 4 & 8 & 16 & \\
        \hline
        BM3D\cite{yang2018bm3dnet}     & 0.798 & 0.853 & 0.867 & 0.885 & 0.851 & 1.218 & 1.186 & 1.169 & 1.154 & 1.182  & 22.415 & 27.556 & 31.997 & 32.936 & 28.726 \\
        DnCNN\cite{zhang2017beyond}    & 0.856 & 0.876 &0.896 & 0.922 & 0.887 & 0.726 & 0.754 & 0.785 & 0.803 & 0.767 & 27.561 & 29.533 & 32.215 & 35.269 & 31.144 \\
        RIDNet\cite{anwar2019real}   &0.857 & 0.875 & 0.886 & 0.902 & 0.880 & 0.705 & 0.734 & 0.765 & 0.809 &0.753 & 28.581 & 30.573 & 33.044 & 35.328 & 31.881 \\
        N2N\cite{lehtinen2018noise2noise}  &0.887 & 0.903 & 0.924 & 0.946 & 0.915 & 0.827 & 0.841 & 0.866 & 0.879 & 0.853 & 30.565 & 34.754 & 36.886 & 38.445 & 35.162 \\
        N2V\cite{krull2019noise2void}    & 0.890 & 0.911 & 0.932 & 0.941 & 0.918 & 0.801 & 0.836 & 0.903 & 0.912 & 0.863 & 29.276 & 30.571 & 31.359 & 32.986 & 31.048 \\
        RDDM\cite{liu2024residual}    & 0.915 & 0.926 & 0.934 & 0.948& 0.931 & 0.851 & 0.873 & 0.883 & 0.889 & 0.874 & 31.438 & 33.453 & 34.469 & 35.556 & 33.729 \\
        \textbf{Ours} & \textbf{0.934} & \textbf{0.941} & \textbf{0.949} & \textbf{0.958} & \textbf{0.946} & \textbf{0.938} & \textbf{0.942} & \textbf{0.953} & \textbf{0.976} & \textbf{0.952} & \textbf{32.305} & \textbf{35.349} & \textbf{38.145} &\textbf{41.561}  & \textbf{36.841} \\
        \hline
    \end{tabular}
    }
    \caption{Quantitative comparisons with other methods in terms of the SSIM, HFRR and PSNR on the W2S dataset\cite{zhou2020w2s}.}
  \label{Tab:2}
\end{table}

\begin{figure*}[htbp]
\begin{center}
 \includegraphics[width=1.0\linewidth]{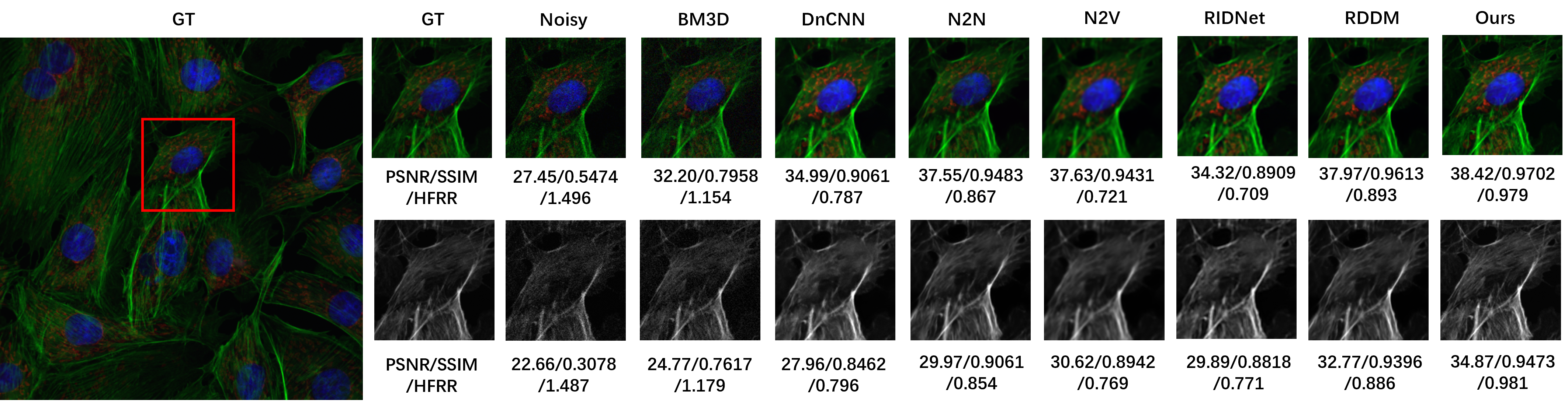}
\end{center}
   \caption{Qualitative comparison on FMD dataset \cite{zhang2019poisson}. Left: Ground-truth images. Right: Magnified views of different image denoising results.(Up: RGB channel Down: Green channel)}
\label{fig:4}
\end{figure*}

\noindent \textbf{Fluorescence Image Noise Removal.} The experimental results on both the FMD and W2S biomedical microscopy datasets comprehensively validate the effectiveness and generalizability of QWD-GAN. On the FMD dataset (Table \ref{Tab:1}), although the diffusion model RDDM slightly outperforms in SSIM under high-noise conditions (0.936 vs. 0.938), QWD-GAN achieves superior performance in PSNR (33.311\,dB vs. 33.452\,dB) and High-Frequency Retention Ratio (HFRR) (0.958 vs. 0.893), indicating a significantly better ability to preserve image details and textures. As further illustrated in Figure \ref{fig:4}, QWD-GAN provides clearer restoration of high-frequency structures such as cell membranes and microfilaments in both RGB and green channels, whereas RDDM tends to suffer from over-smoothing or hallucinated structures. These advantages stem from the synergistic design of the wavelet-guided multi-scale denoising generator (WG-MSDG) and the frequency-spatial feature fusion module (FS-FF), which allow the model to suppress noise while maintaining rich high-frequency biological information. Additionally, the quality-aware dual-branch discriminator enhances perceptual quality and structural consistency by integrating image quality assessment (IQA) features. More notably, on the W2S wide-field microscopy dataset (Table \ref{Tab:2}), QWD-GAN consistently outperforms all baseline methods, achieving the highest PSNR (36.841\,dB), SSIM (0.946), and HFRR (0.952), surpassing RDDM's 33.729\,dB, 0.931, and 0.874, respectively. These results highlight QWD-GAN’s strong adaptability to large-scale structural variations and complex texture distributions. In particular, under the spatially heterogeneous and multi-scale conditions prevalent in W2S, QWD-GAN still accurately reconstructs fine structural details, showcasing its practical value and broad potential in real-world biomedical image denoising tasks.

\begin{figure}[H]
\begin{center}
 \includegraphics[width=0.9\linewidth]{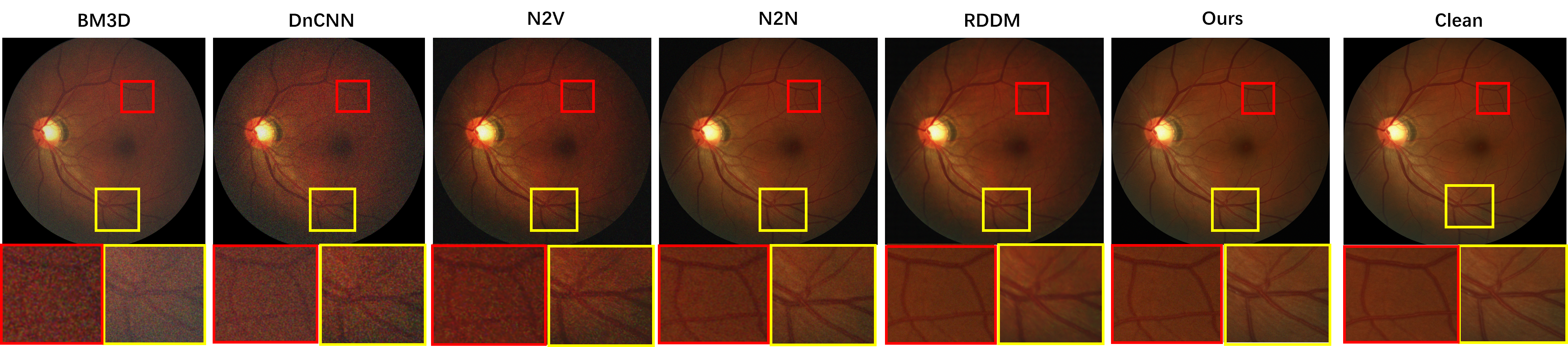}
\end{center}
\vspace{-0.5em}
   \caption{Qualitative results of our method and other baselines on REFUGE \cite{orlando2020refuge}. The display window is yellow [400;400] and red [250;250]. }
\label{fig:5}
\end{figure}

\noindent \textbf{Evaluation on Fundus Dataset.} In this section, we evaluate the generalization ability of the proposed method on real-world medical image denoising, i.e.Fundus dataset REFUGE \cite{orlando2020refuge}. This is a dataset of 1200 fundus images with ground truth segmentation and clinical glaucoma labels, which can be used for retinal disease screening training. To verify the importance of our proposed method for clinical medical image denoising, we randomly selected 300 fundus images for verification, and the denoising performance is shown in Table \ref{Tab:3} and Figure \ref{fig:5}.

\begin{table}[H]
\centering
\resizebox{0.9\columnwidth}{!}{
\begin{tabular}{lcccccc}
\hline
\textbf{Methods} & \textbf{BM3D}\cite{yang2018bm3dnet} & \textbf{DnCNN}\cite{zhang2017beyond} &  \textbf{N2N}\cite{lehtinen2018noise2noise} & \textbf{N2V}\cite{krull2019noise2void} & \textbf{RDDM}\cite{liu2024residual} & \textbf{Ours} \\
\hline
SSIM ↑      & 0.837 & 0.842  & 0.876 & 0.872 & \textbf{0.915} & 0.902 \\
HFRR ($\rightarrow 1$)
      & 1.224 & 0.833  & 0.856 & 0.861 & 0.884 & \textbf{0.946} \\
PSNR ↑ (dB) & 27.924 & 29.269  & 32.454 & 32.986 & 36.556 & \textbf{36.561} \\
\hline
\end{tabular}
}
\vspace{0.5em}
\caption{Average SSIM, HFRR and PSNR results of different methods on the 300 randomly selected from the Fundus dataset REFUGE \cite{orlando2020refuge}.}
\label{Tab:3}
\end{table}

\noindent \textbf{Comparison with Other SOTA Denoising Methods Across Datasets}

\begin{table*}[htbp]
\centering
\resizebox{\textwidth}{!}{%
\begin{tabular}{llccccc}
\hline
\textbf{Category} & \textbf{Method} & \textbf{SIDD Validation \cite{abdelhamed2018high}} & \textbf{SIDD Benchmark \cite{abdelhamed2018high}} & \textbf{FMD \cite{zhang2019poisson}} & \textbf{W2S \cite{zhou2020w2s}} & \textbf{REFUGE \cite{orlando2020refuge}} \\
\hline

\multirow{9}{*}{Zero-shot} 

& WNNM \cite{gu2014wnnm} & 26.05/0.592 & 25.78/0.809 & - & - & - \\
& DIP \cite{ulyanov2018deep} & 32.10/0.740 & 31.97/0.854 & 31.72/0.912 & 25.61/0.812 & 35.61/0.912\\
& Self2Self \cite{quan2020self2self} & 29.46/0.595 & 29.51/0.651 & 30.76/0.695 & 27.52/0.826 & 35.63/0.932 \\
& PD-denoising \cite{zhou2020awgn} & 31.87/0.820 & 33.61/0.890 & 32.86/\textbf{0.913} & 27.62/0.852 & 36.62/0.944 \\
& NN+denoising \cite{zheng2020unsupervised} & 33.18/\textbf{0.899} & 33.10/0.895 & 32.21/0.872 & 27.66/0.860 & 36.52/0.943 \\
& APBSN-single \cite{lee2022ap} & 30.90/0.818 & 30.76/0.815 & 31.47/0.891 & 26.67/0.852 & 35.22/0.931 \\
& ScoreDVI \cite{cheng2023score} & 34.75/0.856 & 34.70/0.890 & 33.10/0.896 & 27.77/\textbf{0.867} & 37.09/0.845 \\
& MASH \cite{chihaoui2024masked} & 35.06/0.851 & 34.89/0.920 & 33.01/0.888 & 27.71/0.853 & 36.87/0.935 \\
& blind-spot denoising  \cite{quan2025zero} & \textbf{35.31}/0.868 & \textbf{35.05/0.922} & \textbf{33.95}/0.885 & \textbf{27.88}/0.859 & \textbf{37.20/0.948} \\
\hline

\multirow{8}{*}{Self-supervised } 

& NBR2NBR \cite{huang2021neighbor2neighbor} & 27.94/0.604 & 27.90/0.679 & - & - & - \\
& CVF-SID \cite{neshatavar2022cvf} & 34.81/0.944 & 34.71/0.917 & 32.73/0.843 & 29.86/0.907 & 33.29/0.913 \\
& LUD-VAE \cite{zheng2022learn} & 34.91/0.944 & 34.70/0.920 & 33.59/0.915 & 30.99/0.905 & 35.48/0.925 \\
& R2R \cite{pang2021recorrupted} & 35.04/0.844 & 34.78/0.898 & - & - & - \\
& APBSN \cite{lee2022ap} & \textbf{36.73}/0.878 & 36.69/0.874 & - & - & - \\
& PUCA \cite{jang2023puca} & - & 37.54/0.936 & - & - & - \\
& SelfFormer \cite{quan2024pseudo} & - & \textbf{37.69}/0.937 & - & - & -
\\
& Ours & 36.31/\textbf{0.956} & 37.65/\textbf{0.942} & \textbf{33.95/0.936} & \textbf{32.30/0.934} & \textbf{36.56/0.946}
\\
\hline
\end{tabular}%
}
\caption{Comparison of different methods (PSNR(dB)/SSIM) on SIDD Validation, SIDD Benchmark, FMD,W2S, and REFUGE datasets. Best results are in \textbf{bold}.}
\label{tab:benchmark_results}
\end{table*}

Table \ref{tab:benchmark_results} presents a comprehensive quantitative comparison between the proposed QWD-GAN and a series of representative state-of-the-art denoising methods across multiple datasets, including SIDD Validation, SIDD Benchmark \cite{abdelhamed2018high}, FMD \cite{zhang2019poisson}, W2S \cite{zhou2020w2s}, and REFUGE  \cite{orlando2020refuge}. The compared approaches cover zero-shot methods (e.g., WNNM \cite{gu2014wnnm}, DIP \cite{ulyanov2018deep}, Self2Self \cite{quan2020self2self}, MASH \cite{chihaoui2024masked}), and self-supervised or unpaired learning methods (e.g., NBR2NBR \cite{huang2021neighbor2neighbor}, CVF-SID \cite{neshatavar2022cvf}, R2R \cite{pang2021recorrupted}, SelfFormer \cite{quan2024pseudo}).

From the results, QWD-GAN consistently achieves the best or near-best performance across all datasets. Specifically, on SIDD Validation and Benchmark, QWD-GAN obtains the highest PSNR/SSIM scores (36.31/0.956 and 37.65/0.942, respectively), demonstrating strong generalization to real-world smartphone imaging noise. On biomedical datasets (FMD and W2S), QWD-GAN shows clear superiority in both PSNR and SSIM, with notable improvements in preserving fine structural details compared to diffusion-based and self-supervised baselines. Furthermore, on the REFUGE dataset, QWD-GAN achieves the highest SSIM (0.946) and PSNR (36.56 dB), highlighting its robustness and clinical applicability in fundus image denoising.

It is worth noting that several entries in Table \ref{tab:benchmark_results} are left blank. This is because not all methods provide results across every dataset or noise condition. In particular, some classical algorithms (e.g., WNNM \cite{gu2014wnnm}, DIP \cite{ulyanov2018deep}) were originally evaluated only on synthetic Gaussian noise and lack reported results on biomedical benchmarks. Similarly, for certain self-supervised frameworks (e.g., R2R \cite{pang2021recorrupted}, PUCA \cite{jang2023puca}), official implementations are not available for all datasets, and retraining them under identical conditions is often computationally prohibitive or infeasible. Therefore, blank cells indicate that no fair or directly comparable results were available, rather than a failure of the method.

\vspace{-0.5em} 

\subsection{Ablation Study}

\noindent \textbf{Quantitative Results.} Table \ref{Tab:4} presents the results of each variant. We report PSNR, SSIM, and wavelet-domain MAE to evaluate image fidelity and frequency-domain consistency. These results collectively demonstrate that each component of our WG-MSDG contributes meaningfully to its denoising performance, both in pixel accuracy and structural fidelity.

\begin{table}[ht]
\centering
\small
\label{tab:ablation}
\begin{tabular}{lccc}
\hline
\textbf{Method} & \textbf{PSNR ↑} & \textbf{SSIM ↑} & \textbf{Wavelet MAE ↓} \\
\hline
Full WG-MSDG (Ours)         & \textbf{32.35} & \textbf{0.934} & \textbf{0.0213} \\
w/o FS-FF                   & 31.96 & 0.881 & 0.0311 \\
w/o Wavelet Branch          & 31.02 & 0.876 & 0.0455 \\
w/o WCG Block               & 28.54 & 0.850 & 0.0387 \\
w/o $\mathcal{L}_{\text{wavelet}}$ & 29.63 & 0.853 & 0.0420 \\
Naive Fusion                & 29.71 & 0.856 & 0.0342 \\
\hline
\end{tabular}
\caption{Ablation study on key components of the proposed WG-MSDG model on W2S dateset by averaging 1 noisy frames.}
\label{Tab:4}
\end{table}

\begin{table}[htbp]
\centering
\small
\setlength{\tabcolsep}{4.5pt}

\label{tab:ablation2}
\begin{tabular}{lcccccccc}
\hline
\textbf{Method} & \textbf{IQA} & \textbf{Fusion} & \(\mathcal{L}_{\text{IQA}}\) & PSNR ↑ & SSIM ↑ & LPIPS ↓ & FID ↓ \\
\hline
Baseline (PatchGAN) & \ding{55} & – & \ding{55} & 31.84 & 0.882 & 0.188 & 38.5 \\
A: +IQA (no fusion) & \ding{51} & – & \ding{55} & 32.12 & 0.895 & 0.165 & 35.4 \\
B: +IQA + Channel    & \ding{51} & Chan. & \ding{55} & 32.58 & 0.802 & 0.142 & 32.7 \\
C: +IQA + Spatial    & \ding{51} & Spatial & \ding{55} & 32.66 & 0.909 & 0.137 & 31.2 \\
D: B + \(\mathcal{L}_{\text{IQA}}\) & \ding{51} & Chan. & \ding{51} & 33.05 & 0.917 & 0.121 & 28.4 \\
E: C + \(\mathcal{L}_{\text{IQA}}\) & \ding{51} & Spatial &\ding{51} & \textbf{33.31} & \textbf{0.938} & \textbf{0.109} & \textbf{26.8} \\
\hline
\end{tabular}
\caption{Ablation study on the quality-aware discriminator. We investigate the contributions of the IQA module, fusion strategy, and IQA loss on the FMD dataset by averaging 1 noisy frames. Best results are \textbf{bolded}.}
\label{Tab:5}
\end{table}

\vspace{-2em} 

\section{Conclusion}

This paper introduces QWD-GAN, an unsupervised denoising framework for biomedical microscopy images that combines a multi-scale wavelet-based generator with a quality-aware dual-branch discriminator. By fusing spatial and frequency features and enforcing wavelet-domain consistency, the model effectively preserves high-frequency details while suppressing noise. Experiments on multiple medical datasets (FMD, W2S, REFUGE) show that QWD-GAN outperforms existing methods in SSIM, PSNR, and HFRR, demonstrating strong generalization and clinical applicability.


\newpage 

\bibliography{main}
 \end{document}